\documentclass[epj]{webofc}
\usepackage[varg]{txfonts}   
\usepackage{epsfig}
\usepackage{graphicx}
\usepackage{mathptmx}      
\usepackage{bm}

\def\tr{{\rm tr}}

\newcommand{\ignore}[1]{}

\wocname{EPJ Web of Conferences}
\woctitle{INPC 2013}
\begin{document}
\title{Fluid/Gravity Correspondence, Second Order Transport and Gravitational Anomaly~\thanks{Talk given by E.~Meg\'{\i}as at the International Nuclear Physics Conference INPC 2013, 2-7 June 2013, Firenze, Italy.} 
\fnsep
\thanks{Supported by Plan Nacional de Altas Energ\'{\i}as (FPA2009-07908, FPA2011-25948), Spanish MICINN Consolider-Ingenio 2010 Programme CPAN (CSD2007-00042), Comunidad de Madrid HEP-HACOS S2009/ESP-1473, Spanish MINECO's Centro de Excelencia Severo Ochoa Program (SEV-2012-0234, SEV-2012-0249), and the Juan de la Cierva Program.}
}
%

\author{Eugenio~Meg\'{\i}as\inst{1}\fnsep\thanks{\email{emegias@ifae.es}} \and
        Francisco~Pena-Benitez\inst{2,3}
}

\institute{Grup de F\'{\i}sica Te\`orica and IFAE, Departament de F\'{\i}sica, Universitat Aut\`onoma de Barcelona, Bellaterra E-08193 Barcelona, Spain
\and
          Instituto de F\'{\i}sica Te\'orica UAM/CSIC, C/ Nicol\'as Cabrera 13-15, Universidad Aut\'onoma de Madrid, Cantoblanco E-28049 Madrid, Spain 
\and
Departamento de F\'isica Te\'orica, Universidad Aut\'onoma de Madrid, Cantoblanco E-28049 Madrid, Spain
          }

\abstract{We study the transport properties of a relativistic fluid affected by chiral and gauge-gravitational anomalies. The computation is performed in the framework of the fluid/gravity correspondence for a 5 dim holographic model with Chern-Simons terms in the action. We find new anomalous and non anomalous transport coefficients, as well as new contributions to the existing ones coming from the mixed gauge-gravitational anomaly. Consequences for the shear waves dispersion relation are analyzed.}
\maketitle

\section{Introduction}
\label{intro}

Hydrodynamics is an useful approach to study many phenomena for physical systems out of equilibrium.  It can be applied when the mean free path of particles is much shorter than the characteristic size of the system~\cite{Kovtun:2012rj}, and it is formulated in terms of the effective field theory formalism. The basic ingredients are the constitutive relations, i.e. expressions of the energy-momentum tensor and the current in terms of fluid quantities (charge density, fluid velocity, etc), organized in a derivative expansion,
\begin{eqnarray}
\langle T^{\mu\nu} \rangle &=& (\epsilon + p) u^\mu u^\nu + p \eta^{\mu\nu} + \tau_{(1)}^{\mu\nu}  + \tau_{(1)\textrm{ano}}^{\mu\nu} + \tau_{(2)}^{\mu\nu} + \tau_{(2)\textrm{ano}}^{\mu\nu} + \cdots \,, \label{eq:fullconstiT}\\
\langle J^\mu \rangle &=& n u^\mu + \nu_{(1)}^{\mu}  + \nu_{(1)\textrm{ano}}^{\mu} + \nu_{(2)}^{\mu} + \nu_{(2)\textrm{ano}}^{\mu}  + \cdots\,. \label{eq:fullconstiJ}\end{eqnarray}
Here we have split the contributions in the equilibrium, first order and second order (anomalous + non anomalous) parts. Dissipative effects like the shear and bulk viscosities have been studied for a long time~\cite{Policastro:2001yc}. During the past few years new transport phenomena induced by chiral anomalies have been discovered, and they proved to play a prominent role in the physics of the quark gluon plasma. In particular the charge separation found at RHIC~\cite{Abelev:2009ac} can be explained by the chiral magnetic effect,  in which an extermal magnetic field in the fluid~$B^\mu = \epsilon^{\mu\nu\rho\lambda} u_\nu \partial_\rho A_\lambda$ induces an electric current parallel to the magnetic field~\cite{Fukushima:2008xe}. By the same way a vortex in the fluid~$\omega^\mu = \epsilon^{\mu\nu\rho\lambda} u_\nu \partial_\rho u_\lambda$ induces also an electric current parallel to the vorticity vector, the so-called chiral vortical effect~\cite{Erdmenger:2008rm,Son:2009tf}. Up to this point, only pure gauge anomalies had been considered to be relevant for these effects, but very recently it has been pointed out that mixed gauge-gravitational anomalies contribute also even at first order~\cite{Landsteiner:2011cp,Landsteiner:2011iq,Landsteiner:2012kd}.  In this paper we will sketch the computation of the transport coefficients up to second order, for a strongly coupled CFT with holographic dual in four dimensions, and present some results.

\section{Holographic model and constitutive relations}
\label{sec:holographic_model}

We consider a holographic Einstein-Maxwell model in 5 dim, supplemented with gauge and mixed gauge-gravitational Chern-Simons terms, which realizes a single chiral $U(1)$ symmetry~\cite{Landsteiner:2011iq,Landsteiner:2011tg,Chapman:2012my}
\begin{eqnarray}
&&\hspace{-1cm} S = \frac{1}{16\pi G} \int d^5x \sqrt{-g} \bigg[ R + 2 \Lambda -
  \frac 1 4 F_{MN} F^{MN} + \epsilon^{MNPQR} A_M
  \left( \frac\kappa 3 F_{NP} F_{QR} + \lambda R^A\,_{BNP} R^B\,_{AQR}
  \right) \bigg]   \label{eq:EMmodel} \\
&&\hspace{-0.4cm}+  S_{\textrm{GH}}  +  S_{\textrm{CSK}} \,. \nonumber
\end{eqnarray}
The boundary term $S_{\textrm{CSK}}$ is needed to reproduce the
gravitational anomaly at general
hypersurface~\cite{Landsteiner:2011iq,Landsteiner:2012dm}. The covariant $U(1)$ current is $(16\pi G) J^\mu =  -\sqrt{-\gamma} F^{r\mu} |_\epsilon$. Its divergence leads to the anomaly for chiral fermions~\cite{Bertlmann:1996xk}, and this we use to fix the parameters $\kappa=-G/(2\pi)$ and $\lambda= - G/(48\pi)$. 

Our goal is to compute the transport coefficients up to second order in the constitutive relations, Eqs. (\ref{eq:fullconstiT})-(\ref{eq:fullconstiJ}). We use the Landau frame, defined as the one in which the energy flux vanishes at rest, $u_\mu \tau^{\mu\nu}_{(n)}=0$ (see~\cite{Megias:2013xla} for a discussion of other frames). The most general contributions at first order~are
\begin{equation}
\tau_{(1)}^{\mu\nu}  = -2\eta\sigma^{\mu\nu}\,, \quad  \tau_{(1)\textrm{ano}}^{\mu\nu}  =  0 \,,  \quad\nu_{(1)}^{\mu}  =  -\sigma\left(T P^{\mu\nu}\mathcal{D}_\nu\bar{\mu}-E^\mu\right)\,, \quad  \nu_{(1)\textrm{ano}}^{\mu} =  \xi_B B^\mu +  \xi_V\omega^\mu\,, \label{eq:tau_nu_1}
\end{equation}
where $\mathcal{D}_\nu$ is the Weyl covariant derivative, $P^{\mu\nu} = u^\mu u^\nu + \eta^{\mu\nu}$, $\sigma^{\mu\nu} = \frac{1}{2}(\mathcal{D}^\mu u^\nu + \mathcal{D}^\nu u^\mu)$ and $\bar{\mu} = \mu/T$. In these expressions $\eta$, $\sigma$, $\xi_B$ and $\xi_V$ are the shear viscosity, electrical conductivity, chiral magnetic and vortical conductivities respectively. When including external electromagnetic fields, at second order there are 23 coefficients in the constitutive relation for $\langle T^{\mu\nu}\rangle$, and 15 for $\langle J^\mu \rangle$. A classification of these terms was obtained in~\cite{Kharzeev:2011ds}. We will compute some of them within the present~model.

\section{Fluid/gravity correspondence}
\label{sec:fluid_gravity}

The equations of motion of the model admit an AdS Reissner-Nordstr\"om black-brane solution as long as the parameters of the theory (mass and charge of the black hole, fluid velocity, etc)  are independent of the space-time coordinates $x^\mu$. However, if one assumes that these parameters are slow  varying functions of $x^\mu$, then one can find a solution for the metric and gauge field valid order by order in a derivative expansion~\cite{Erdmenger:2008rm,Banerjee:2008th}. In order to follow this technique, we use the Weyl covariant ansatz,
\begin{eqnarray}
ds^2 &=& -2W_1(\rho)u_\mu dx^\mu \left(dr+r\mathcal{A}_\nu dx^\nu\right) + r^2\left[W_2(\rho)\eta_{\mu\nu}+W_3(\rho)u_\mu u_\nu + 2\frac{W_{4\sigma}(\rho)}{r_+}  P^\sigma_{\mu}u_\nu \right. \label{eq:ds2} \\
&& \left. +\frac{W_{5\mu\nu}(\rho)}{r_+^2}\right] dx^\mu dx^\nu \,,\\
A &=& \left(a^{(b)}_\mu+ a_\nu(\rho)P^\nu_\mu  + r_+ c(\rho) u_\mu\right)dx^\mu \,, \label{eq:A}
\end{eqnarray}
where $\mathcal{A}_\nu =u^\mu D_\nu u_\mu -\frac{1}{3}D_\mu u^\mu$ is the Weyl connection, $r_+$ is the outer horizon of the black hole and $\rho = r/r_+$. This solution leads to the following energy-momentum tensor and current,
\begin{eqnarray}
&&\hspace{-1.2cm}\langle T_{\mu\nu} \rangle  = \frac{ 1}{16\pi G}\lim_{\epsilon\to 0}\left( r_+^4(W_2+W_3)^{(\bar{4},\epsilon)}( 4u_\mu u_\nu +\eta_{\mu\nu} ) + 4r_+^2W_{5\mu\nu}^{(\bar{4},\epsilon)} + 8r_+^3W^{(\bar{4},\epsilon)}_{4\sigma} P^\sigma_{(\mu} u_{\nu)} + T^{\textrm{ct}}_{\mu\nu}\right) \,, \label{eq:Treno} \\
&&\hspace{-1.0cm}\langle J_\mu \rangle  = \frac{1}{8\pi G}\lim_{\epsilon\to 0} \left(r^3_+c^{(\bar{2},\epsilon)}u_\mu+ r_+^2 a_\mu^{(\bar{2},\epsilon)} + J^{\textrm{ct}}_\mu\right) \,, \label{eq:Jreno}
\end{eqnarray}
where $F^{(\bar{m},\epsilon)}$ denotes the coefficient of the term $(\rho^{-1}-\epsilon)^{m}$ in an expansion around the regularized boundary with cut-off surface $\rho = 1/\epsilon$.  Because we are considering a flat background metric, the divergences appear only through terms involving electromagnetic fields  in addition to the cosmological constant~\cite{Landsteiner:2011iq,Megias:2013joa}. So we need counterterms, $T^{\textrm{ct}}_{\mu\nu}$ and $J^{\textrm{ct}}_\mu$, to make the expressions finite. 

Inserting this ansatz into the Einstein-Maxwell equations ($E_{AB}= 0$, $M_A = 0$) we find a set of 14 differential equations and 5 constraints. The fields can be decomposed in $SO(3)$ scalar, vector and tensor components. We find a constraint in the vector sector related to the energy conservation, $\mathcal{D}_\mu T^\mu_i=F_{i\alpha}J^\alpha$, and two (one) dynamical equations in the vector (tensor) sector, which read
\begin{eqnarray}
\mathbb J^{(n)}_{i}(\rho)  &=& \partial_\rho\left(\rho^5\partial_\rho W^{(n)}_{4i}(\rho) + 2\sqrt{3}Q a^{(n)}_i(\rho)\right)\,, \label{eq:J_vector} \\
\mathbb  A^{(n)}_{i}(\rho) &=& \partial_\rho\left(\rho^3f(\rho)\partial_\rho a^{(n)}_{i}(\rho) +2\sqrt{3}Q\partial_\rho W^{(n)}_{4i}(\rho)\right) \,,  \label{eq:A_vector} \\
\mathbb P^{(n)}_{ij}(\rho) &=& \partial_\rho\left(\rho^5f(\rho)\partial_\rho W^{(n)}_{5ij}(\rho)\right) \,, \label{eq:P_tensor}
\end{eqnarray}
corresponding to $E_{ri} = 0$, $M_i = 0$ and $E_{ij}-\frac{1}{3}\delta_{ij}\tr(E_{kl}) = 0 $ respectively. While the form of the homogeneous part of these equations is the same at any order, the source terms $\mathbb J^{(n)}_{i}$, $\mathbb A^{(n)}_{i}$ and $\mathbb P^{(n)}_{ij}$ are specific for each order. Their expressions are rather complicated at second order, and we will present here only the sources at first order
\begin{eqnarray}
\mathbb J^{(1)}_{\mu}(\rho) &=&  \lambda\frac{96}{\rho^5}\left( M\rho^2 - 5 Q^2 \right)\frac{B_\mu}{r_+} + \lambda \frac{16\sqrt{3} Q}{\rho^7}\left(  20 M \rho^2- 63 Q^2 \right) \omega_\mu \,, \\
\nonumber\mathbb A^{(1)}_{\mu}(\rho) &=& - \frac{\sqrt{3} \pi  T}{M r_+ \rho ^2}P_\mu^{\nu}\mathcal{D}_\nu Q  - \left(1+\frac{9 Q^2}{2 M \rho ^2}\right)\frac{E_\mu}{r_+} - \kappa\frac{16\sqrt{3} Q}{\rho^3}  \frac{B_\mu}{r_+}  - \kappa \frac{48Q^2}{\rho^5} \omega_\mu \\
&&-\lambda \frac{48}{\rho^{11}} \left(4 M^2 \rho ^4 - 16 M Q^2\rho ^2 + 15 Q^4\right) \omega_\mu  \,, \\
\mathbb P^{(1)}_{\mu\nu}(\rho) &=& -6r_+\rho^2\sigma_{\mu\nu}  \,,
\end{eqnarray}
where the mass $M$ and the charge $Q$ of the black hole are related to the temperature and chemical potential as $T = r_+(2M - 3Q^2)/(2\pi)$ and $\mu = \sqrt{3}r_+Q$. By solving Eqs.~(\ref{eq:J_vector})-(\ref{eq:P_tensor}) with these sources and the appropriate boundary conditions, we can compute the functions $W^{(1)}_{4i}(\rho)$, $ a^{(1)}_i(\rho)$ and $W^{(1)}_{5ij}(\rho)$. Then from their near boundary expansion and using  Eqs.~(\ref{eq:Treno})-(\ref{eq:Jreno}) we obtain the constitutive relations.\footnote{The parameter $c^{(\bar{2},\epsilon)}$ in Eq.~(\ref{eq:Jreno}) can always be chosen to be zero, as it just redefines the charge and mass of the black hole. In addition, in the Landau frame there is no contribution to $\langle T^{\mu\nu}\rangle$ coming from the scalar sector and $(W_2 + W_3)^{(\bar{4},\epsilon)}=0$.}

\section{Transport coefficients}
\label{sec:transport_coefficients}

Using previous technology, we get the first order transport coefficients for the holographic plasma
\begin{eqnarray}
\eta &=& \frac{r_+^3}{16\pi G} \,, \qquad \sigma =\frac{\pi T^2}{16 G M^2 r_+} \,,\\
\xi_B &=& -\frac{\sqrt{3} Q \left(M+3\right)r_+ \kappa }{8 \pi G M} + \frac{\sqrt{3} \pi Q T^2 \lambda }{G M r_+} \,, \qquad  \xi_V = -\frac{3 Q^2 r_+^2 \kappa }{4 \pi G M} + \frac{2 \pi  \left(2 Q^2 - 1 \right) T^2 \lambda }{G M} \,,
\end{eqnarray}
which enter in Eq.~(\ref{eq:tau_nu_1}). One can identify in these expressions the anomalous conductivities $\xi_B$ and $\xi_V$, which are induced by the chiral~$\sim \kappa$ and gauge-gravitational~$\sim\lambda$ anomalies. At second order we present here the result for only $3$ out of the $38$ coefficients (see~\cite{Megias:2013joa} for a computation of all of them)
\begin{equation}
\tau^{\mu\nu}_{(2)\textrm{ano}} = \tilde\Lambda_1 \mathcal{D}^{\langle\mu}\omega^{\nu\rangle}  + \tilde\Lambda_4   \mathcal{D}^{\langle\mu} B^{\nu\rangle} + \cdots  \,, \qquad \nu^\mu_{(2)\textrm{ano}} =  \tilde\xi_5 \epsilon^{\mu\nu\alpha\beta}u_\nu \mathcal{D}_\alpha E_\beta  + \cdots \,,
\end{equation} 
where $X^{\langle\mu\nu\rangle}$ stands for transverse traceless projection. $\tilde\Lambda_1$, $\tilde\Lambda_4$ and $\tilde\xi_5$ are dissipative as they are odd under time reversal. As a consequence, they can be naturally factorized in terms of relaxation lenghts as $\tilde{\Lambda}_1 = -2\eta\tilde{l}_\omega$, $\tilde{\Lambda}_4 = -2\eta\tilde{l}_B$ and $\tilde{\xi}_5 = \sigma\tilde{l}_E$, in analogy to the relaxation time $\tau_{\pi}$~\cite{Israel:1979wp}. Our result is
\begin{eqnarray}
\tilde{l}_\omega &=& \frac{2\pi}{ G  p}\left[ \frac{\kappa  \mu ^3}{48\pi^2   } +64\mu\lambda\left( 3 r_+^2 - 2\mu^2 - \frac{\pi  T \mu ^2}{r_+}\right) \right] \,,  \label{eq:l_omega}\\
\tilde{l}_B &=& \frac{1}{2\pi G p}\left( \frac{\kappa  \mu ^2}{8} +  \lambda \pi^2 T^2  \right) \,, \qquad \tilde{l}_E =  -\frac{8 \bar\mu }{\pi^2 T}\left[\kappa  \log 2 - 2 \lambda  (1+2\log 2)\right]+  \mathcal{O}(\bar{\mu}^3)\,. \label{eq:le}
\end{eqnarray}
It is remarkable that $\tilde{\xi}_5$ is not vanishing in our model, even when it was predicted to be zero based on no entropy production arguments~\cite{Kharzeev:2011ds}. We have cross-checked this result with a computation using Kubo formulae. In particular, when switching on a gauge field $A_y = A_y(t,z)$, the coefficient~$\tilde{\xi}_5$ appears in the two point function $\left\langle J^x(k) J^y(-k) \right\rangle = -i\xi_B k_z + \sigma \tilde l_E \omega k_z  + \cdots$. This Kubo formula leads to the result~(\ref{eq:le}) even in the probe limit. Another observation comes from the dispersion relation of shear waves. In~\cite{Kharzeev:2011ds} they have found~$\omega \approx -i\eta/(4p)k^2 \mp i C k^3\,,$ with $C = -\tilde{\Lambda}_1/(8p)$. Our computation leads to $C =  \tilde l_\omega \eta/(4p) $. The chiral anomalous contribution $\sim\kappa$ of Eq.~(\ref{eq:l_omega}) agrees with the result in~\cite{Kharzeev:2011ds}. In addition, we found a new anomalous contribution induced by the gauge-gravitational anomaly $\sim\lambda$.

\section{Conclusions}
\label{sec:conclusions}

We have studied the transport properties of a holographic model that implements the chiral and gauge-gravitational anomalies. We found new gravitational anomaly contributions at second order appearing in $13$ coefficients in the constitutive relation for $\langle T^{\mu\nu}\rangle$, and in $11$ coefficients for $\langle J^\mu\rangle$. In addition, new coefficients involving electromagnetic fields have been computed for the first time. A derivation of the entropy current~\cite{Chapman:2012my} up to second order within the present model will be very interesting as it would give some insights on the role of some anomalous processes, like~$ \langle J^\mu \rangle \sim \tilde\xi_5 \epsilon^{\mu\nu\alpha\beta}u_\nu \mathcal{D}_\alpha E_\beta$.

%
%
%

\end{document}